\documentclass[12pt,preprint,iop]{emulateapj}

\usepackage{epsfig, natbib} 
\usepackage{mathrsfs,amssymb}
\usepackage{graphicx,latexsym}

\newcommand{\etal}{{\rm et~al.\/}}

\newcommand{\mum}{\mbox{$\mu \rm m$}}

\slugcomment{Submitted to ApJ 2013 June 24, accepted December 15 2014, published March 25 2015}

\shorttitle{SAFIRES}
\shortauthors{Hanish \etal}

\begin{document}

\title{The Spitzer Archival Far-InfraRed Extragalactic Survey}

\author{D.J.\ Hanish\altaffilmark{1},
        P.\ Capak\altaffilmark{1},
        H.I. Teplitz \altaffilmark{1},
	V.\ Desai\altaffilmark{1},
	L.\ Armus \altaffilmark{1},
	C.\ Brinkworth \altaffilmark{1},
        T.\ Brooke\altaffilmark{1},
	J.\ Colbert \altaffilmark{1},
	L.\ Edwards \altaffilmark{2},
	D.\ Fadda\altaffilmark{1},
        D.\ Frayer\altaffilmark{3},
	M.\ Huynh \altaffilmark{4},
	M.\ Lacy \altaffilmark{5},
	E.\ Murphy \altaffilmark{6},
	A.\ Noriega-Crespo \altaffilmark{1},
	R.\ Paladini \altaffilmark{1},
	C.\ Scarlata \altaffilmark{7},
	S.\ Shenoy \altaffilmark{8}}

\altaffiltext{1}{Spitzer Science Center, California Institute of Technology, MC 220-6, 1200 E California Blvd., Pasadena, CA 91125, $danish@alumni.caltech.edu$}
\altaffiltext{2}{Astronomy Department, 260 Whitney Avenue, Yale University, New Haven, CT 06511}
\altaffiltext{3}{National Radio Astronomy Observatory, P.O. Box 2, Green Bank, WV 24944}
\altaffiltext{4}{International Centre for Radio Astronomy Research, M468, University of Western Australia, Crawley, WA 6009, Australia}
\altaffiltext{5}{National Radio Astronomy Observatory, 520 Edgemont Road, Charlottesville, VA 22903}
\altaffiltext{6}{The Observatories of the Carnegie Institution for Science, CA 91101}
\altaffiltext{7}{Minnesota Institute for Astrophysics, School of Physics and Astronomy, University of Minnesota, Minneapolis, MN 55455}
\altaffiltext{8}{Space Science Division, NASA Ames Research Center, M/S 245-6, Moffett Field, CA 94035}

\begin{abstract}

We present the Spitzer Archival Far-InfraRed Extragalactic Survey (SAFIRES).  This program produces refined mosaics and source lists for all far-infrared extragalactic data taken during the more than six years of the cryogenic operation of the $Spitzer\ Space\ Telescope$.  The SAFIRES products consist of far-infrared data in two wavelength bands (70 \mum\ and 160 \mum) across approximately 180 square degrees of sky, with source lists containing far-infrared fluxes for almost 40,000 extragalactic point sources.  Thus, SAFIRES provides a large, robust archival far-infrared data set suitable for many scientific goals.

\end{abstract}

\section{Introduction} \label{s:intro}

The $Spitzer\ Space\ Telescope$, launched in 2003, has produced a wealth of infrared data products across the mid- and far-infrared (FIR) regimes \citep{b:werner04}.  The $Spitzer$ imaging instruments during the cryogenic mission were the Infrared Array Camera \citep[IRAC,][]{b:fazio04} and the Multiband Imaging Photometer \citep[MIPS,][]{b:rieke04}.  The MIPS instrument ceased operations when the telescope's cryogen supply expired in 2009, leaving a rich legacy of FIR data.  These FIR bands have proven to be useful for a number of scientific purposes, such as the modeling of interstellar gas and dust, characterization of quasar energy distributions, and constraining the star formation properties of the most luminous infrared galaxies.

The Spitzer Archival FIR Extragalactic Survey (SAFIRES) is an offshoot of the $Spitzer\ Space\ Telescope$ Enhanced Imaging Products (SEIP) \citep{b:capak13}.  The SEIP produces reliable, consistent data products for five mid-infrared $Spitzer$ bands, including all four IRAC bands (3.6, 4.5, 5.8, and 8.0 \mum) as well as the 24 \mum\ channel of MIPS.  The use of the entire cryogenic $Spitzer$ sample results in unequaled depth of coverage in many well-studied fields, along with a robust source list containing as many as fifty million sources across an area of over 1,500 square degrees.  This source list is then linked to counterparts in the WISE and 2MASS source catalogs, extending the range of wavelengths further.

SAFIRES applies the SEIP project's methods to the remaining two MIPS bands, located at far-infrared wavelengths of 70 and 160 \mum.  Due to the complexity of far-infrared observations, these bands require an expansion of SEIP's standard pipeline through the addition of reprocessing tools.  These additional steps are required to remove obvious artifacts before extracting useful measurements.  As a result, these bands were not included in the SEIP project, but were later funded through an additional NASA Astrophysics Data Analysis Program (ADAP) grant.  To ensure high reliability, the SAFIRES sample includes no fields near the Galactic disk; these observations comprised more than half of the area observed by $Spitzer$, but the practical drawbacks of Galactic contamination would inhibit the ability to maintain the level of reliability desired in the SAFIRES products.  As with SEIP, SAFIRES' source list contains no extended sources.  The remaining sample comprises nearly two thousand fields spanning almost two hundred square degrees of sky.

In this paper we present the SAFIRES data products, available online from the NASA/IPAC Infrared Science Archive (IRSA), hosted by the Infrared Processing and Analysis Center at the California Institute of Technology.\footnote{http://irsa.ipac.caltech.edu}  Section~\ref{s:fields} describes the fields comprising the SAFIRES sample.  Section~\ref{s:mosaics} explains the process used to create the SAFIRES ``supermosaics''.  Section~\ref{s:sourcelists} describes how the official lists of high-confidence sources were generated.  Section~\ref{s:conc} summarizes the SAFIRES results.

\section{Fields}\label{s:fields}

{ 
\begin{figure*}[!t]
\centerline{\hbox{\includegraphics[width=10.5cm]{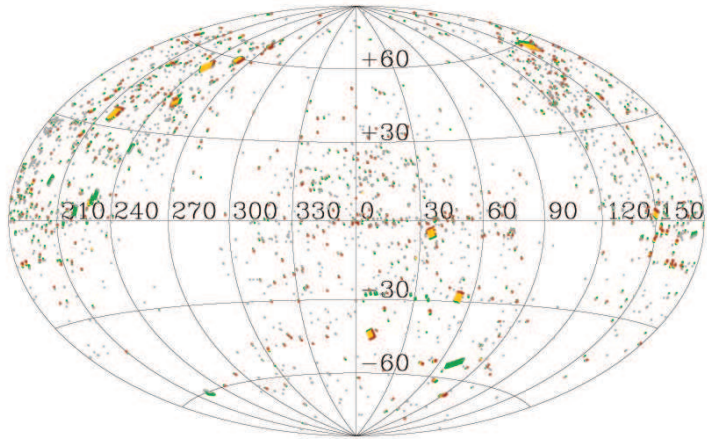}}}
\caption[Coverage]{Areas covered by the two SAFIRES bands, in equatorial coordinates.  Areas covered in both bands are shown in yellow, those with only 70 \mum\ MIPS data are in green, and those with only 160 \mum\ data are in red.  Grey areas have 24 \mum\ MIPS data in the SEIP, but no SAFIRES data.  Sizes of small regions are enhanced for greater visibility.}
\label{f:coverage}
\end{figure*}}

The SAFIRES sample consists of all extragalactic ($\vert$b$\vert \ge 20.0\deg$) fields observed by the $Spitzer$ Space Telescope during six years of cryogenic operation for which 70 \mum\ or 160 \mum\ data is available.  Many of these observations were never used as part of any published work, and a significant number of additional fields benefitted from an increased depth due to overlapping observations from multiple projects.  As many larger fields were observed on multiple occasions, this stacking often results in a depth of coverage far superior to that used by the original requesting teams.

The extragalactic criterion used for the SAFIRES sample was due primarily to the difficulties involved in correcting for the effects of large-scale contamination from bright objects, such as foreground stars in our galaxy.  Due to the ways in which the SAFIRES image processing scripts reduce the artifacts within each image's background, Galactic contamination can greatly impact far-infrared flux measurements in unpredictable ways.  While this limitation removes as much as half of the usable $Spitzer$ area from consideration, limiting the SAFIRES sample to extragalactic fields maintains the high reliability seen in the SEIP source lists.

The final SAFIRES data products include 1132 distinct regions with data in either far-IR band.  The majority of these (1106 small regions) each consist of a single contiguous field with area less than 1.0 deg$^2$.  Many of these small regions were pointed observations of specific bright objects, such as various types of AGN, and contain little else. Others comprise parallel fields not targeted on specific objects; the MIPS instrument could obtain data in multiple channels simultaneously, but with an angular separation on the sky between them.  These incidental FIR data were taken in the course of obtaining 24 \mum\ observations of nearby discrete objects.  Many small SAFIRES fields thus consist of two linked areas separated by a region unobserved in each far-infrared band.

The remaining twenty-six SAFIRES regions were each larger than 1.0 deg$^2$ in at least one $Spitzer$ band.  In addition to possessing more extensive coverage areas, these largest fields include many of the most well-studied deep areas, such as GOODS-N and GOODS-S \citep{b:giavalisco04}, the Lockman Hole \citep{b:lockman86}, and COSMOS \citep{b:scoville07}.  The superior depths of coverage of these fields result in a noticeably higher density of detected sources than is seen in the smaller, more isolated fields, and the variety of characteristics in these sources is much higher.  Information about these large fields is given in Table~\ref{t:bigfields}, and comparable information for the smaller fields is available online.

In total, 177.93 square degrees of extragalactic fields are covered by at least one of the two far-IR MIPS bands; 66.53 deg$^2$ have only 70 \mum\ data, 23.80 deg$^2$ have only 160 \mum, and 87.60 deg$^2$ have both.  Nearly all of this area overlaps with the MIPS 24 \mum\ band, with only 30.13 deg$^2$ having 70/160 \mum\ data but no corresponding 24 \mum\ observations; the majority of these areas consist of parallel 70/160 \mum\ fields taken during 24 \mum\ observations.  The distributions of fields within each band are shown in Figure~\ref{f:coverage}.

{ 
\begin{deluxetable*}{l | c c | c c | c c | c c | c}
  \tablewidth{0pt}
  \tabletypesize{\small}
  \tablecaption{SAFIRES supermosaics \label{t:bigfields}}
  \tablehead{\colhead{Supermosaic} &
             \colhead{RA} &
	     \colhead{Dec} &
             \multicolumn{2}{c}{Area [deg$^2$]} &
             \multicolumn{2}{c}{Sensitivity$^{\dag}$} &
	     \multicolumn{2}{c}{Number of sources} &
	     \colhead{Field} \\
	     \colhead{ID} &
	     \multicolumn{2}{c}{(J2000) [deg]} &
             \colhead{70 \mum} &
             \colhead{160 \mum} &
             \colhead{70 \mum} &
             \colhead{160 \mum} &
             \colhead{70\mum} &
             \colhead{160\mum} &
             \colhead{name} }
\startdata
60075851 & 202.38 & ~11.80 & ~3.43 & ~0.01 & ~29.41 & ~~5.35 & 176 & 0 & --- \\
60076221 & 172.80 & ~-3.98 & ~0.76 & ~0.01 & ~35.07 & ~~9.47 & 31 & 0 & --- \\
60077621 & 204.04 & ~38.30 & ~0.70 & ~0.72 & ~~6.44 & ~27.79 & 260 & 156 & --- \\
60081181 & ~36.76 & ~~0.74 & ~0.41 & ~0.40 & ~16.39 & ~53.07 & 44 & 50 &  ---\\
60082211 & ~~8.83 & -43.58 & ~7.62 & ~7.68 & ~14.17 & ~61.11 & 1216 & 744 & ELAIS-S1 \\
60083651 & 202.08 & ~33.66 & ~0.61 & ~0.67 & ~13.85 & ~67.03 & 92 & 26 & --- \\
60086021 & 249.32 & ~40.91 & ~5.71 & ~5.66 & ~~8.52 & ~44.61 & 1746 & 837 & ELAIS-N2 \\
60087831 & 217.80 & ~34.28 & 12.11 & 12.04 & ~~8.56 & ~35.50 & 3900 & 2441 & SDWFS/ELAIS-N3 \\
60087881 & 229.40 & ~~7.44 & ~8.17 & ~1.45 & ~27.07 & ~20.84 & 861 & 242 & --- \\
60088431 & 154.27 & ~38.85 & ~1.46 & ~1.49 & ~12.23 & ~55.88 & 344 & 286 & --- \\
60091101 & 246.64 & ~42.41 & ~0.89 & ~1.31 & ~~7.29 & ~35.66 & 218 & 88 & --- \\
60095751 & 259.14 & ~59.56 & ~5.65 & ~5.49 & ~10.32 & ~57.88 & 1825 & 890 & --- \\
60095761 & 161.33 & ~58.06 & 12.55 & 12.41 & ~10.17 & ~38.65 & 3533 & 2254 & Lockman Hole \\
60095791 & 242.96 & ~54.90 & 10.52 & 10.44 & ~~9.31 & ~42.25 & 2953 & 1793 & ELAIS-N1 \\
60095801 & 207.13 & ~26.43 & ~1.00 & ~1.07 & ~18.46 & ~60.29 & 164 & 50 & --- \\
60095821 & ~67.14 & -61.31 & ~1.01 & ~1.06 & ~13.68 & ~73.88 & 214 & 62 & --- \\
60095871 & ~35.56 & ~-4.46 & 10.37 & 10.35 & ~16.74 & ~58.90 & 1934 & 1472 & XMM-LSS \\
60095881 & ~52.83 & -28.56 & ~8.81 & ~8.86 & ~~9.89 & ~41.84 & 2372 & 1595 & CDFS/GOODS-S \\
60095931 & 150.27 & ~~2.21 & ~4.21 & ~4.20 & ~20.14 & ~71.29 & 672 & 508 & COSMOS \\
60096041 & ~~9.80 & -27.38 & ~1.01 & (N/A) & ~19.87 & ~(N/A) & 88 & (N/A) & --- \\ 
60096191 & 222.95 & -60.28 & ~3.49 & (N/A) & 346.50 & ~(N/A) & 224 & (N/A) & --- \\
60096201 & ~90.95 & -66.56 & ~0.87 & ~1.15 & ~20.46 & 166.22 & 75 & 0 & --- \\
60096211 & 269.76 & ~66.63 & ~0.86 & ~1.19 & ~20.22 & 150.12 & 73 & 0 & --- \\
60096291 & 220.29 & ~~3.48 & ~3.46 & (N/A) & ~33.45 & ~(N/A) & 133 & (N/A) & --- \\
60096301 & 241.34 & ~23.80 & ~3.46 & (N/A) & ~23.53 & ~(N/A) & 168 & (N/A) & --- \\
60096371 & ~71.11 & -53.74 & 11.44 & (N/A) & ~~8.97 & ~(N/A) & 1486 & (N/A) & S-SEP \\
70101860 & 189.34 & ~62.30 & ~0.69 & ~0.75 & ~~4.57 & ~21.07 & 226 & 120 & GOODS-N \\
70101880 & 214.60 & ~52.83 & ~1.10 & ~1.20 & ~~4.64 & ~22.59 & 434 & 154 & EGS
\enddata
\tablecomments{
\dag: Sensitivities are in mJy/beam for a 10$\sigma$ detection. \\
The fields listed are those 26 contiguous areas large enough in 70 or 160 \mum\ to necessitate separation into multiple discrete mosaics.  A similar table giving information for the 1106 additional small SAFIRES fields is available online.}
\end{deluxetable*}}

\section{Mosaics} \label{s:mosaics}

The SAFIRES mosaics were produced using a method similar to that of the overall SEIP source list pipeline outlined in \citet{b:capak13}.  Contiguous areas on the sky were identified using the footprints of the IRAC and MIPS 24 \mum\ images.  Images for contiguous areas smaller than 2 square degrees were merged into discrete ``supermosaic'' images through the MOsaicker and Point source EXtractor \citep[][MOPEX]{b:makovoz04}.  Areas larger than 2 deg$^2$ were further split into a series of overlapping tiles, each with a central region 30 arcminutes square with an additional 5\'\ buffer on each side to ensure all individual exposures affecting the central region are included in the stack.  Each contiguous region was assigned a unique 8-digit ``region ID'' (REGID) number, and individual tiles within the larger supermosaics are indicated with a hyphenated subregion number.

In all, 26 large fields possessing SAFIRES data were separated into a total of 813 0.25 deg$^2$ images,  with the largest fields containing nearly one hundred individual mosaics per wavelength band.  An additional 1106 small contiguous areas possessed MIPS 70 or 160 \mum\ data.  While SAFIRES added MIPS 70 and 160 \mum\ data to the image pipeline, there were no areas possessing data in these two bands that did not also have IRAC or MIPS 24 \mum\ data, and so the addition of these two bands did not affect the region designations.  The final product therefore includes 1919 individual subregions, each possessing mosaics in at least one of the two SAFIRES bands as well as at least one of the five SEIP bands.

The two far-infrared MIPS bands possess larger background variations than are common in the more spatially resolved bands, requiring extra processing before accurate measurements can be made.  This is due to a number of factors, such as the low number of absolute pixels contained in each far-infrared image and the sheer size of the point spread function (PSF) in the MIPS 70 and 160 \mum\ bands.  The most significant source of error is caused by contamination from bright or extended objects.  These sources will often create negative ``side-lobes'' above and below bright objects, as well as affecting the intensity of the object itself, while extended sources can find their flux canceled out almost entirely.  These erroneous features will have a substantial impact on both source detection and flux measurement, and so must be mitigated or removed wherever possible.


SAFIRES expands SEIP's MOPEX-based pipeline through the use of the Germanium Reprocessing Tools (GeRT), a toolkit designed for the reduction of MIPS 70 and 160 \mum\ data\footnote{MOPEX and GeRT are available from IRSA, along with their documentation}.  Details of the principles and algorithms used to analyze MIPS data can be found in \citet{b:gordon04}.  The GeRT package helps account for background contamination through the use of column and high-pass filters, designed to remove these irregularities in all far-infrared MIPS images. Our processing follows the procedure outlined in \citet{b:frayer09}, but we will briefly explain the process as it applies to SAFIRES below. 

The $cleanup70.tcsh$ and $cleanup160.tcsh$ scripts within GeRT use a user-supplied list of point sources in each image, mask the regions around each of these bright objects, and perform a time median filter on the data to create corrected versions of the input BCD images.   Specifically, the provided positions of bright objects are used to create a mask for use in a column fit algorithm, removing circular areas of fixed radii centered around the input positions to improve the quality of the background gradient estimation.   The corrected BCD images can then be used in the original MOPEX mosaic generation script to produce mosaics with reduced background corruption from nearby bright objects.

The SAFIRES processing scripts use the Astronomy Point source EXtractor (APEX) package within MOPEX \citep{b:makovoz05} to locate the brightest objects within the corresponding MIPS 24 \mum\ mosaic for each region (if available), using conservative criteria designed to identify bright point sources capable of corrupting the output mosaic.  These positions are then fed into the aforementioned GeRT scripts to produce mosaics with reduced contamination from bright objects.  The mask radii used by this process can be adjusted by the user, but only one size can be used for all point sources in a given image, and larger sizes result in more uncertain image backgrounds as the number of masked pixels rises.  To account for this, the default radius used for this process is slightly larger than those of a typical point source, but is insufficient to account for extended objects.  As a result, objects more extended than the mask or bright enough that the PSF wings extend beyond the mask are not completely removed and will continue to create artifacts, specifically negative lobes; the intensities of these lobes will be substantially reduced because the central regions of the objects are removed.  Due to this residual effect, the SAFIRES source list does not include extended objects (see Section~\ref{s:sourcelists}).  These artifacts are especially pronounced for bright 160 \mum\ objects near the edges of a mosaic and we recommend studies attempting to analyze these particular 160 \mum\ sources should perform their own GeRT reprocessing to improve these images further.


Once the GeRT scripts corrected the input images, the standard SEIP image pipeline was then used to produce our final data products.  The first step in this process used the MOPEX ``overlap'' pipeline to match backgrounds across individual GeRT-modified exposures.  The IDL-based AFRL overlap algorithm was used for mosaics were the standard MOPEX overlap method failed to converge, as well as for those few areas possessing a number of exposures that would make the standard algorithm computationally infeasible.  The exact algorithm used in each mosaic is recorded within the image header.  After background matching, the MOPEX ``mosaic'' pipeline was run to produce a combined mosaic.  This pipeline first rejects cosmic rays and other image artifacts using a combination of the ``dual-outlier'', ``outlier'', and ``box-outlier'' packages as appropriate.  Next, the images within each contiguous area were combined into an exposure time weighted mean image as well as a median image.  In addition to the mosaic, this process also generated an exposure coverage map, mean uncertainty image, median uncertainty image, and a map of the standard deviation of the inputs at each pixel position.

\section{Source Lists} \label{s:sourcelists}

{ 
\begin{deluxetable*}{c | c | c}
  \tablewidth{0pt}
  \tabletypesize{\footnotesize}
  \tablecaption{SAFIRES source list format \label{t:ourlist}}
  \tablehead{\colhead{Column} &
             \colhead{Name} &
             \colhead{Descriptions} }
\startdata
(1) & objID & Source ID name within the SAFIRES database \\
(2) & SMID & Supermosaic ID \\
(3) & REGID & Region ID from SEIP; see \citet{b:capak13} \\
(4)--(5) & RA, DEC & Right Ascension and Declination (J2000) \\
(6) & ID & Source ID number within each mosaic \\
(7)--(8) & PIX\_X, PIX\_Y & Source position, in pixels, within each mosaic \\
(9)--(10) & PSFflux, dPSFflux & Flux and uncertainty of a PSF centered on the listed pixel \\
(11) & FWHM & Full width at half maximum for the detected object \\
(12) & Coverage & Depth of coverage at the object's central position \\
(13) & SNR & Signal-to-noise ratio of detected object \\
(14)--(15) & F\_AP18p0, dF\_AP18p0 & Flux and uncertainty for a fixed 18.0$\tt''$ circular aperture centered at the listed pixel$^{\dag}$ \\
(16) & Area & Area of detected object \\
(17) & Gain & Gain of detected object \\
(18) & MeanNoise & Mean noise level of the mosaic containing detected object
\enddata
\tablecomments{Source lists are available online for the two far-infrared SAFIRES bands.  These lists consist of the above information for all point sources meeting the SAFIRES minima for coverage and SNR. \\
Note that all values are band-specific; separate source catalogs are available for each of the two SAFIRES bands. \\
\dag: For 70 \mum\ images the fixed aperture is of radius 18.0$\tt''$; for 160 \mum\ the aperture is 40$\tt''$, with a corresponding change in column name.}
\end{deluxetable*}}

{ 
\begin{figure*}[!t]
\centerline{\hbox{\includegraphics[width=7.5cm]{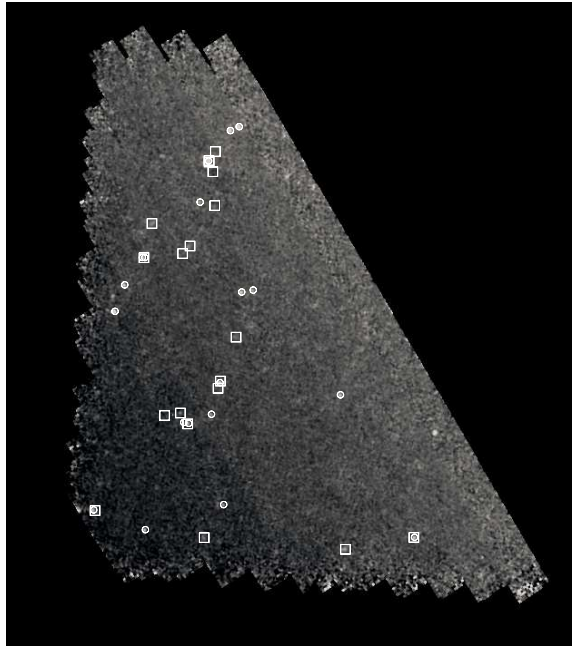}}}
\caption[Object Detections]{Detected 70 \mum\ and 160 \mum\ sources in a sample area, region ID 60095761-68.  The image shown is the SAFIRES 70 \mum\ mosaic; 70 \mum\ detections are shown with small circles, with 160 \mum\ using larger squares.}
\label{f:detects}
\end{figure*}}

The SAFIRES project includes the generation of robust lists of compact sources in the MIPS 70 \mum\ and 160 \mum\ bands.  The mechanism for generating source lists in these bands is very similar to that described in \citet{b:capak13} for MIPS 24 \mum\ data; using the APEX package within MOPEX with a 10$\sigma$ threshold, we identify individual extragalactic point sources within each field.  As with SEIP, the SAFIRES source lists are solely made up of point sources to guarantee confidence in the resulting source catalogs.  While extended objects can be measured in far-infrared fields, the large measurement apertures would add unacceptable levels of uncertainty to the measured fluxes of all but the brightest extended objects.  Additionally, the background-fitting methods of GeRT were designed for point sources; extended objects will be handled incorrectly by the reprocessing tools, and so will suffer from both negative features near the object, as well as a partial removal of legitimate extended flux.  As a result, the methods used to generate the SAFIRES source lists are only reliable for compact sources; given the large and complex PSFs of the two SAFIRES bands, and the tendency for multiple sources to blend, this criterion removes many possible detections from these source lists.  Extended sources are identified with Source Extractor, using significantly modified parameter sets.  While a number of parameters are modified to detect extended sources, the primary SAFIRES requirement for these objects is that they possess 120 or more pixels with flux values more than 10 standard deviations above the sky background.

Figure~\ref{f:detects} shows the limits of this process; while the majority of bright objects are identified in the final source list, several identifiable objects were rejected by our various selection criteria; as with the SEIP catalogs, the SAFIRES source lists favor confidence over completeness.  The basic format of the provided source lists is shown in Table~\ref{t:ourlist}.  While most of the variables are self-explanatory, the flux variables are worth mentioning in more detail.  The SAFIRES and SEIP source lists include fluxes derived using PSF fitting methods (PSFflux) as well as fluxes derived from fixed, band-specific apertures (F\_AP18p0 for 70 \mum\ data).  Depending on the specific science goals or characteristics of the observed object, a user may prefer a corrected aperture flux, although the PSF fit is recommended as SAFIRES includes no Galactic or extended sources where a PSF fit algorithm would be more likely to introduce significant errors.

{ 
\begin{figure*}[!t]
\centerline{\hbox{\includegraphics[width=10.5cm]{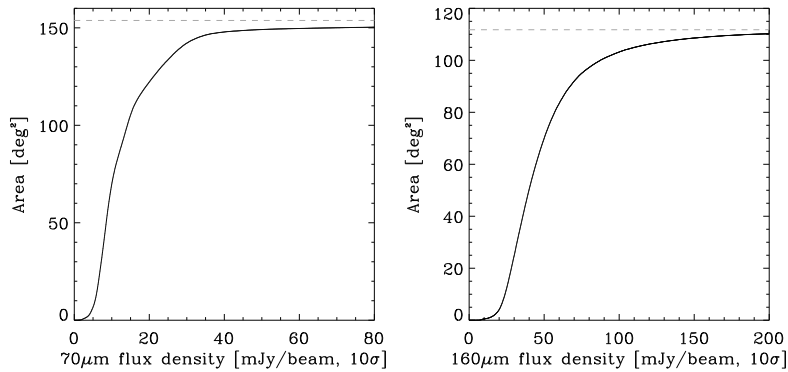}}}
\caption[Cumulative histograms]{Cumulative histograms showing total areas for which an object with a given flux density could be detected with 10$\sigma$ sensitivity, for all fields observed in the two SAFIRES bands.}
\label{f:depthhist}
\end{figure*}}

Unlike the source lists generated by SEIP, we make no attempt to link the sources found in the two SAFIRES bands to each other, or to any sources in the SEIP bands.  The primary reason for this is source blending; the diameter of the PSF for a source in the 160 \mum\ band is approximately 40 arcseconds, which nearly always means that multiple mid-IR objects are located within the FIR PSF envelope.  For consistently luminous point sources, such as the type 1 quasars of \citet{b:hanish13}, this effect can safely be ignored as the bright source will far outshine any of its incidental neighbors.  As the SAFIRES source catalogs are intended to be more general than that, however, we can make no similar assumptions for the sample as a whole.

{ 
\begin{figure*}[!t]
\centerline{\hbox{\includegraphics[width=10.5cm]{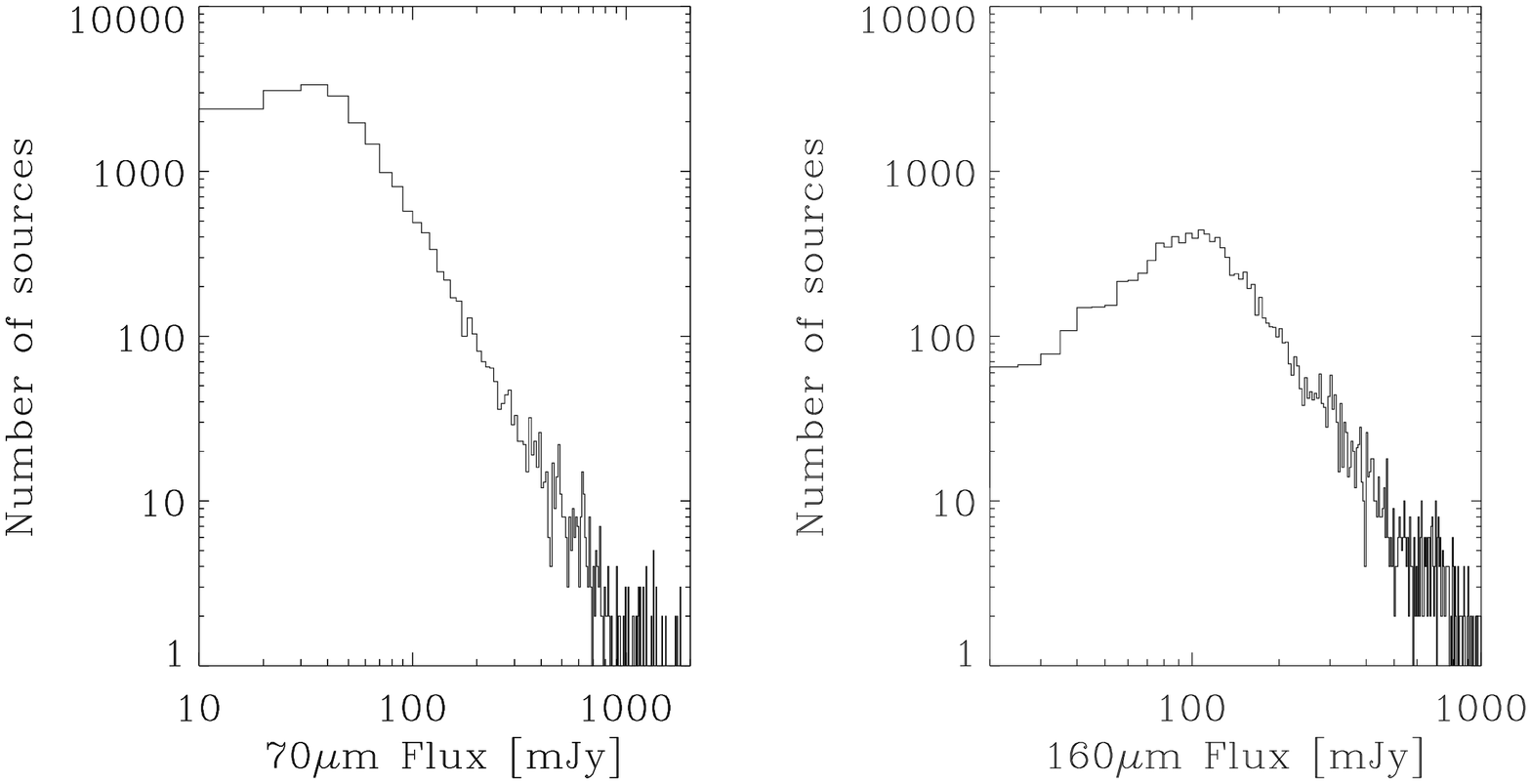}}}
\caption[Source fluxes]{Histograms of fluxes of 10$\sigma$ sources within all SAFIRES fields within the two FIR MIPS bands.}
\label{f:fluxhist}
\end{figure*}}

Note that the majority of point sources simply do not exceed a 10$\sigma$ detection threshold in the far-infrared MIPS bands, due to the variations in background levels.  This is especially problematic for fields with relatively low depths of coverage, which includes many fields aimed at specific classes of target.  While a lower SNR threshold would give a number of detections more directly comparable to those in the more resolved SEIP bands, we prefer to maintain the higher cutoff to ensure the reliability of these source lists.
Users wishing to determine fluxes for specific known objects outside of the deep, well-studied fields are encouraged to download the SAFIRES mosaics and perform aperture measurements around positions derived from more resolved bands.  This method was used by \citet{b:hanish13}, drawing quasar positions from SDSS, although this method is not encouraged for sources not likely to outshine any potential blended neighbors.

The SAFIRES source lists contain a total of 27523 sources in the 70 \mum\ band, as well as 14612 in the 160 \mum\ band.  Over 90\% of source detections in each SAFIRES band were found in the largest, deepest fields.  The methods used to generate the source lists for SAFIRES emphasize reliability and confidence, but these lists are not complete source catalogs containing all objects within each field.  Figure~\ref{f:depthhist} shows the sensitivities of the SAFIRES fields; while the majority of the individual fields are relatively shallow, the far greater area covered by the few extremely deep fields results in typical infrared point sources being easily detectable within the majority of the SAFIRES area.  Figure~\ref{f:fluxhist} shows the distribution of fluxes for the point sources included within the source lists; as expected, the majority of sources in the SAFIRES source lists have flux values well below the 10$\sigma$ sensitivity of the shallower fields, and so would not be detected in those areas.

{ 
\begin{figure*}[!t]
\centerline{\hbox{\includegraphics[width=10.5cm]{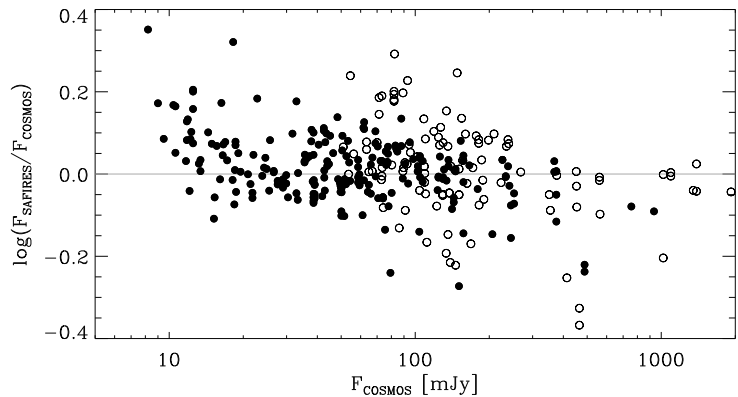}}}
\caption[COSMOS comparison]{Comparison of SAFIRES PSF-derived fluxes to those drawn from the COSMOS $Spitzer$ source lists of \citet{b:frayer09}.  Solid symbols denote 70 \mum\ measurements, with open circles for 160 \mum.}
\label{f:cosmoscomp}
\end{figure*}}

We compare our fluxes to those from the previous COSMOS work of \citet{b:frayer09}, as shown in Figure~\ref{f:cosmoscomp}.  Both works use similar methods, performing a PSF-based flux measurement on MIPS 70 and 160 \mum\ mosaics through the APEX software package.  The data sets used are also similar; SAFIRES mosaics are somewhat deeper, due to the inclusion of additional overlapping observations, and the SAFIRES COSMOS-including region (60095931) covers almost twice as much total area as the original COSMOS field.  This results in substantially better depth along what had been the fringes of the original COSMOS observations.

As can be seen in Figure~\ref{f:cosmoscomp}, the majority of SAFIRES fluxes agree with those of \citet{b:frayer09} with a standard deviation of approximately 10\%.  In addition to the expected scatter due to low flux, the much larger variations at the faint end of each band are almost entirely caused by two issues.  The majority of the points are located in regions with relatively low depth of coverage, often near the edges of the original COSMOS field.  Additionally, one of the major differences between this work and \citet{b:frayer09} is the manner in which blended sources are treated; many of the drastically differing sources are treated as multiple objects in one work and a single object in the other.  Nevertheless, the SAFIRES fluxes appear to agree well with those of previous $Spitzer$ studies.

\section{Conclusions} \label{s:conc}

The SAFIRES project has expanded the previous $Spitzer$ imaging products to include the majority of extragalactic observations taken over six years of cryogenic operation.  These data have been combined into a series of mosaic images, from which reliable lists of point sources have been generated.  These archival data products can be used for many scientific purposes in lieu of direct observations, either directly or with minimal processing by prospective users.

The final SAFIRES data products contain mosaics spanning approximately 180 deg$^2$ of extragalactic sky, containing a mix of small pointed observations and large, deep surveys.  The provided source lists for SAFIRES include 27523 sources in the MIPS 70 \mum\ band and 14612 in the 160 \mum\ band.  Due to the large PSF for these bands, it is possible that each of these sources contains flux contributed by multiple objects, especially in the 160 \mum\ MIPS band; these fluxes should generally be assumed to be an upper limit, although for the extremely bright objects the effects of blending are minimal.  No attempt is made to correlate these catalogs to each other, or to the mid-infrared source lists of SEIP.  Nevertheless, the SAFIRES source lists contain reliable far-infrared fluxes for many of the brightest objects, useful for a variety of scientific purposes.  For sources below the detection thresholds of the single-band methods, the combination of the SAFIRES mosaics with positions derived from more spatially resolved bands allow for even more accurate multiwavelength correlation.
\\
\acknowledgments

Support for this work was provided by NASA through contract NASA-ADAP NNX10AD52G.  This publication makes use of raw data from the Spitzer Space Telescope, operated by the Jet Propulsion Laboratory/California Institute of Technology under NASA contract, as well as data products from the Two Micron All Sky Survey, a joint project of the University of Massachusetts and the Infrared Processing and Analysis Center/California Institute of Technology, funded by NASA and the NSF.


\end{document}